\documentclass[11pt]{article}
\usepackage{moriond,epsfig}
\usepackage{latexsym}

\def\Journal#1#2#3#4{{#1} {\bf #2}, #3 (#4)}

\def\apj{ApJ}					
\def\apjl{ApJ}					
\def\aap{A\&A}					
\def\mnras{MNRAS}				

\def\be{\begin{equation}}
\def\ee{\end{equation}}
\def\bea{\begin{eqnarray}}
\def\eea{\end{eqnarray}}

\def\hpc{$h^{-1}$Mpc }

\def\z{$\,${\it z}$\,$}
\def\etal{{\it et al}}
\def\ie{{\it i.e.\ }}
\def\del{\delta}                

\begin{document}
\vspace*{4cm}
\title{WITNESSING THE BUILD-UP OF THE COLOUR--DENSITY RELATION}

\author{O. CUCCIATI$^{1,2}$, A. IOVINO$^{1}$, C. MARINONI$^{3}$ and the VVDS Collaboration\footnote{
O. Ilbert, 	  
S. Bardelli, 	  
P. Franzetti, 	  
O. Le F\`evre, 	  
A. Pollo, 	  
G. Zamorani, 	  
A. Cappi,	    
L. Guzzo,	    
H.J. McCracken, 	  
B. Meneux, 	  
R. Scaramella, 	  
M. Scodeggio,		
L. Tresse,		
E. Zucca,		
D. Bottini,		
B. Garilli,		
V. Le Brun,		
D. Maccagni,		
J.P. Picat,		
G. Vettolani,		
A. Zanichelli, 	  
C. Adami, 	  
M. Arnaboldi, 	  
S. Arnouts, 	  
M. Bolzonella, 	  
S. Charlot, 	  
P. Ciliegi, 	  
T. Contini, 	  
S. Foucaud, 	  
I. Gavignaud, 	  
B. Marano, 	  
A. Mazure, 	  
R. Merighi, 	  
S. Paltani, 	  
R. Pell\`o, 	  
L. Pozzetti,	       
M. Radovich,	       
M. Bondi,	       
A. Bongiorno, 	  
G. Busarello, 	  
S. de la Torre, 	  
L. Gregorini, 	  
F. Lamareille, 	  
G. Mathez, 	  
Y. Mellier, 	  
P. Merluzzi, 	  
V. Ripepi,	     
D. Rizzo,	     
S. Temporin, 	  
D. Vergani}}

\address{${}^{1}$INAF-Osservatorio Astronomico di Brera, Via Brera 28, \\ 
20121, Milan, Italy\\
${}^{2}$Universit\'a di Milano-Bicocca, Dipartimento di Fisica, \\
Piazza della Scienza, 3, 20126 Milan, Italy\\
${}^{3}$Centre de Physique Th\'eorique, UMR 6207 CNRS-Universit\'e de Provence, \\ 
F-13288 Marseille, France}

\maketitle\abstracts{
We investigate the redshift and luminosity evolution of the
galaxy colour-density relation using the data from the First Epoch
VIMOS-VLT Deep Survey (VVDS) on scales of $R=5$\hpc up to
redshift $\z \sim 1.5$.  While at lower
redshift we confirm the existence of a steep colour-density relation,
with the fraction of the reddest(/bluest) galaxies of the same
luminosity increasing(/decreasing) as a function of density, this
trend progressively disappears in the highest redshift bins
investigated.Our results suggest the existence of an epoch
(more remote for brighter galaxies) characterized by the
absence of the colour-density relation on the $R=5$\hpc scales
investigated.
The rest frame $u^{*}-g'$ colour-magnitude diagram shows a
bimodal pattern in both low and high density environments up to
redshift $z\sim 1.5$. We find that the bimodal distribution is not
universal but strongly depends upon environment.  Both the
colour-density and colour-magnitude-density relations, on the $R=5$\hpc
scales, appear to be a transient, cumulative product of
genetic and environmental factors that have been operating over at
least a period of 9 Gyr.  These findings support an evolutionary
scenario in which star formation/gas depletion processes are
accelerated in more luminous objects and in high density environments:
star formation activity is progressively shifting with cosmic time
towards lower luminosity galaxies (downsizing), and out of high
density environments.}

\section{Introduction}

There is a well known connection between galaxy properties and the 
environment wherein galaxies reside.  These
correlations extend smoothly over a wide range of density
enhancements, from the extreme environment of rich clusters to very
low densities, well beyond the region where the cluster environment is
expected to have much influence.

A key question that still needs to be addressed is whether these
environmental dependencies were established early on when galaxies
first assembled (the so-called `nature' hypothesis), or whether 
they are the present day cumulative end product of multiple processes 
operating over a Hubble time (density-driven evolution, the so-called  
`nurture' scenario). 

A promising approach to addressing these issues involves extending
observations beyond the local universe. Large and deep redshift 
surveys of the universe are the best available
instrument to select a representative sample of the galaxy population
over a broad and continuous range of densities and cosmic
epochs. In this study, we use the VIMOS VLT Deep Survey, 
the largest ($6582$ objects), deepest 
($0.25<z<1.5$), purely-magnitude selected ($I_{AB}\leq 24$)
redshift sample currently available, to explore the colour-density
relation as a function of both luminosity and cosmic time.
In particular the main goal of this investigation 
is to portray the colour-density relation 
at different epochs and evaluate eventual changes in its overall
{\it normalisation} (Butcher \& Oemler effect, Butcher \& 
Oemler~\cite{beo}) and {\it slope} (Dressler effect, 
Dressler~\cite{dressler1980} ).

A more complete description of this work can be found in Cucciati 
\etal~\cite{cucciati2006}.

\section{The Data and the Environment Reconstruction Scheme}\label{data}

The VIMOS VLT Deep Survey (VVDS) in the VVDS-02h field 
has been conceived as a purely flux-limited ($17.5\leq I \leq24$) 
survey (Le F\`evre \etal~\cite{lefevre2005a}).
The first-epoch VVDS-02h data sample extends over a sky area of
0.7$\times$0.7 $deg^2$ ($\sim$ 37$\times$37 \hpc at \z=1.5), and has a
median redshift \z$\sim$0.76.  It contains 5882 galaxies with
secure redshifts with $0.25 \leq z \leq 1.5$. We adopt the cosmology 
$\Omega_m=0.3$, $\Omega_{\Lambda}=0.7$, 
$h=H_{0}/100$. All magnitudes are quoted in the AB system. 

We characterized the environment surrounding 
a given galaxy, at comoving
position {\bf r}, by means of the dimensionless 3D density contrast $\del({\bf r},R)$
smoothed with a Gaussian filter over a typical dimension $R$: $\del({\bf r},R) = 
[\rho({\bf r},R)-\overline{\rho}({\bf r})] / \overline{\rho}({\bf r})$. 
Densities are estimated using appropriate weights to correct for various survey 
observational 
characteristics (sample selection function, target sampling rate, spectroscopic 
success rate and angular sampling rate). Underestimates of $\delta$ 
due to the presence of edges have been corrected dividing measured densities by 
the fraction of the volume of the filter contained within the survey borders.  

To compute densities we exploit all the galaxies in our flux-limited sample: there 
are advantages and drawbacks in this method, therefore as a complementary approach 
we have reconstructed the density field using also a volume-limited subsample
of galaxies ($(M_B-5 \log h) \leq -20.0$). As long as the two approaches suffer 
from different limitations, obtaining consistent results with both of them
allows us to derive more robust conclusions.

We have paid particular attention to calibrate our density
reconstruction scheme: we used simulated mock catalogues 
extracted from GalICS (Hatton \etal~\cite{Hatton2003}) to 
determine the redshift ranges and smoothing length
scales R over which our environmental estimator is not affected by the
specific VVDS observational constraints. 
We conclude that we reliably reproduced the underlying {\it real}
galaxy environment on scales $R\ge 5$\hpc out to z=1.5.

\section{Results}

In this section we present our results on the dependence of galaxy
colours from local density, luminosity and redshift. For this 
analysis we use rest-frame ($u^{*}-g'$) colours,
uncorrected for dust absorption, derived from rest-frame AB absolute
magnitudes (see Ilbert \etal~\cite{ilbert2005}) as computed in the 
$u^{*}$ and $g'$ CFHTLS-MEGACAM photometric system.

\subsection{The colour-density relation: redshift and luminosity dependence}

We explored the combined dependence of the colour-density
relation on redshift and luminosity. To this purpose, we selected
different samples of galaxies, using as luminosity thresholds the
values $(M_B-5 \log h) \leq -19.0, -19.5, -20.0, -20.5, -21.0$
respectively. For each of these samples the fractions of the reddest
($(u^{*}-g') \geq 1.10$) and bluest ($(u^{*}-g') \leq 0.55$) galaxies are shown
in Fig. \ref{figure2} as a function of $\delta$ in four different
redshift bins. As specified in section \ref{data}, we computed local 
densities also using a volume limited sample ($(M_B-5 \log h) \leq -20.0$). 
The results obtained with this recipe show the same
trends as in Fig. \ref{figure2}, although a bit noisier (see Fig. 
\ref{figure2bis}).

Fig. \ref{figure2} shows that not only the
colour segregation weakens as a function of redshift for galaxies of
similar luminosity, but, at a fixed redshift, it strongly depends on
luminosity: for progressively brighter galaxies the colour--density
relationship, as we know it in the local universe, appears at earlier
cosmic times.

To quantify the statistical significance of our findings, we fitted
the points plotted in Fig. \ref{figure2} with a linear relation ($f = a +
b\delta$, where $f$ is the fraction of red or blue galaxies).  
Fig. \ref{figure2bis} shows the slopes $b$ and the associated 1
$\sigma$ error bars as a function of redshift, for red (triangles) 
and blue (squares) galaxies, for
the three subsamples limited at $(M_B-5 \log h) = -19.0, -20.0, -21.0$
going from top to bottom. Left panel refers to Fig. \ref{figure2},
\ie when density contrast is estimated using the full flux limited sample,
while right panel refers to results obtained when density contrast is 
estimated using the volume limited sample. Arrows
indicate the redshift bin where the colour--density relation, as
we know it in the local universe, appears for the first time.

\subsection{The colour-magnitude diagram: redshift and density dependence} 

We also explored the evolution of the distribution of
galaxies in the colour-magnitude plane ($u^{*}-g'$) vs. $(M_B-5 \log h)$
as a function of both redshift and environment.

In Fig. \ref{figure4} the first 3 columns show the isodensity
contours of the distribution of galaxies in different redshift ranges
(from top to bottom as indicated on the right) and for different
environments (from left to right as indicated on top).  The difference
between the over-dense and under-dense colour-magnitude distributions
is shown in the $4^{th}$ column.

The $1^{st}$ column shows that the bimodal distribution of galaxies in
colour space, well established in the local universe (e.g. Strateva \etal 
~\cite{Strateva2001}), persists out to the highest redshift investigated
($z\sim 1.5$). This analysis confirms and extends at higher redshifts
previous results obtained with photometric redshifts out to $z=1$
(e.g. Bell \etal~\cite{Bell2004}, Nuijten \etal~\cite{nuijten2005}). 
The $2^{nd}$ and $3^{rd}$ columns of Fig. \ref{figure4} show that
bimodality holds irrespective of environment out to $z\sim 1.5$.

We can discriminate finer environmental dependencies imprinted in the
bimodal colour distribution by plotting the difference between the
over- and under-dense colour-magnitude diagrams.  The $4^{th}$
column of Fig. \ref{figure4} shows that {\it the colour-magnitude
distribution is not universal but strongly depends upon environment.}  
At low redshift, and for any luminosity, there is a prominent excess of
red objects in over-dense regions, while under-dense regions are
mostly populated by blue galaxies.  On the other hand, and most
interestingly, moving towards higher redshifts the relative ratio of
the two peaks of the bimodal distribution becomes mostly insensitive
to environment (at $0.9<z<1.2$) with the hint of the development of a
more pronounced peak of blue galaxies in high density regions in the
last redshift bin ($1.2<z<1.5$).

\section{Discussion}

\begin{figure}[h]
\centering \psfig{figure=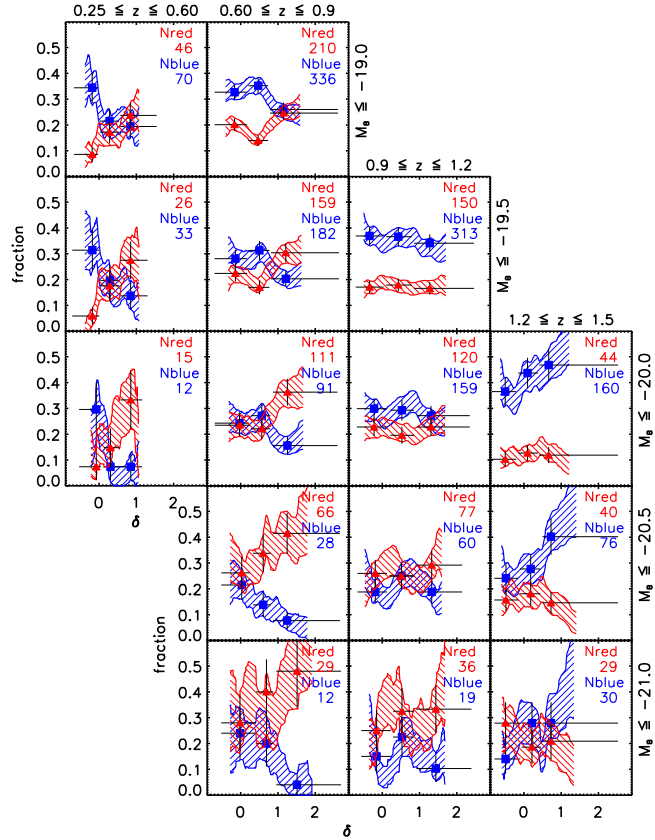,height=4.5in}
\caption{Fraction of the reddest ($(u^{*}-g') \geq 1.10$, triangles)
 and bluest ($(u^{*}-g') \leq 0.55$, squares) galaxies as a
 function of the density contrast $\delta$ in different redshift
 bins (columns) and for different
 luminosity limits (rows).}
 \label{figure2}
\end{figure}

\begin{figure}[h]
\centering \psfig{figure=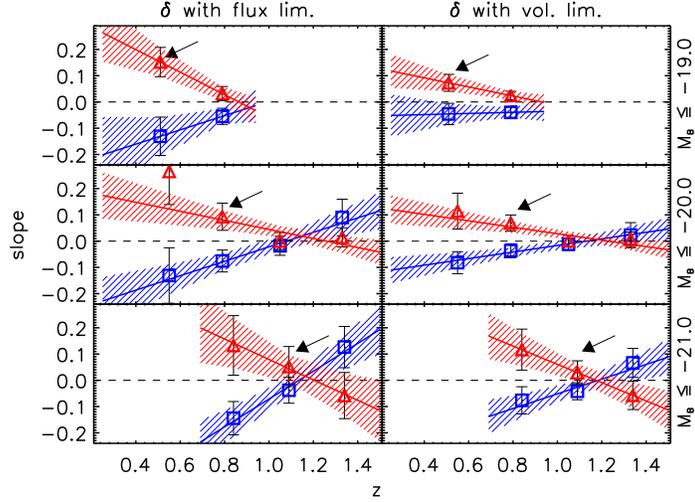,height=2.7in}
\caption{Best fit slopes (and their
associated 1 $\sigma$ error bars) of the fraction of reddest (triangles) 
and bluest (squares) 
galaxies as a function of $\delta$ for the four different redshift
bins of Fig. \ref{figure2} (left column) and for the same results obtained 
when $\delta$ is computed with a volume limited sample (right column). From top to bottom 
different limits in
absolute magnitude are considered, as indicated on the right. The 
straight lines (and the shaded error bar area associated) are the result of the 
linear fit to the points shown.  Arrows indicate the redshift bin where the
colour--density relationship, as we know it in the local universe,
appears for the first time.}
 \label{figure2bis}
\end{figure}

\begin{figure}[h]
\centering \psfig{figure=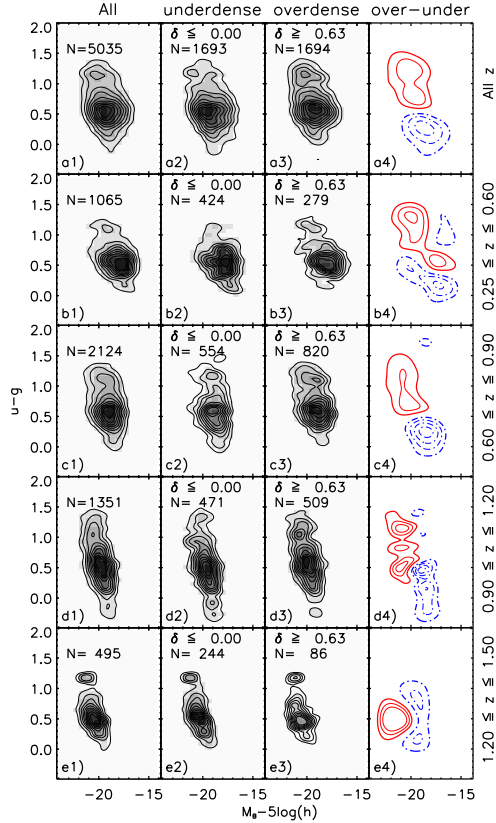,height=4.4in}
\caption{The first 3 columns show the isodensity contours of the
distribution of galaxies in the ($u^{*}-g'$) vs. $(M_B-5\log h)$ plane: 
rows are different redshift ranges (labels on the
right) and columns are different environments (labels 
on top).  The grey scale is normalised to the total
number of objects contained in each panel.  The difference between the
over-dense and under-dense colour-magnitude distributions is shown in
the $4^{th}$ column.  $1-2-3-\sigma$ levels of significance in the
difference are shown using red continuous lines (positive
differences), and blue dotted lines (negative differences). The
thicker lines correspond to $1\sigma$ level.}
 \label{figure4}
\end{figure}

Our study represents the first attempt to use a purely flux-limited redshift
survey to explore the primordial ($\z=1.5$) appearance of the colour-density and
colour-magnitude diagrams from the densest peaks of the galaxy distribution 
down to very poor environments and faint magnitudes, on $R=5$\hpc scales.
Our findings  can be summarised as follows:

a) The most striking result of this study is displayed in Fig.
\ref{figure2}: {\it The colour-density relation shows a dramatic change as a
function of cosmic time}. While at the lowest redshifts explored we confirm 
the existence of a strong
colour-density relation, with the fraction of the red(/blue) galaxies
increasing(/decreasing) as a function of density, at previous epochs 
blue and red galaxies seem to be mostly insensitive
to the surrounding environment, and at the remotest epochs explored ($z
\sim 1.5$) even the most luminous red galaxies do not reside
preferentially in high density environments. 
At $z\sim 1$ there is evidence of absence of the color-density 
relation for medium luminosity galaxies. Moreover, there are hints  
that the well established local trend, which progressively disappears,
eventually reverses in the highest redshift bins investigated ($\sim 1\sigma$ effect),
suggesting that in remote look-back times the star
formation activity was higher in higher density peaks, a property 
reminiscent of a similar characteristic of
Ly-break galaxies (Foucaud \etal~\cite{foucaud2003}).
The absence of the color-density relation  
at the highest redshift bins investigated implies that {\it quenching of star formation was
more efficient in high density regions}.

b) {\it The evolution of the colour-density relation depends on
luminosity}.  Not only, at fixed luminosity, there is a progressive
decrease of red objects as a function of redshift in high densities, 
but also, at fixed redshift, there is a progressive decrease
of fainter red galaxies.  This result implies that {\it star formation
ends at earlier cosmic time for more luminous/massive galaxies}.

c) Not only the slope of the colour-density has changed, but also its
overall normalisation: the relative fraction of the bluest objects was higher in the past
in both high and low density regions and for both brighter
and fainter galaxies. However, the observed drop in the star formation
rate of blue objects in poor environments is weaker than in high
density regions and is weaker for fainter galaxies.  This 
indicates that the mechanisms driving galaxy formation and evolution operate
with different timescales in different environments, and that star formation 
rate continues to be substantial at the
present day in field galaxies, especially in the fainter ones.

We can interpret these findings by making some simplifying
hypothesis. Let's assume, to first order, that the adopted colour
classes are a proxy for different star formation histories, bluer
galaxies having experienced relatively recent star formation. In this
case, the observed time dependence of the colour-density
relation implies that star formation is differentially suppressed in
high and low density regions. For galaxies of similar luminosity the
drop in star formation rate occurred earlier in higher density
environments, resulting in the red excess observed at present epoch,
and progressively later in lower density regions, \ie in the
field, where a larger blue component is still observed. This 
suggests that some environment driven mechanism may be at work. The
drop in star formation is also a function of luminosity (and therefore
probably mass): truncation mechanisms are more efficient in brighter
systems than in fainter ones.

From an observational side our analysis well agrees with the so called
{\it downsizing} scenario, first suggested by Cowie \etal~\cite{cowie96}, but
modified to take into account the observed environmental
dependence. According to our observations, star formation activity is
not only progressively shifted to smaller systems, but also from
higher to lower density environments.

This result agrees remarkably well with our findings (obtained with
the same sample) about the significant evolution of galaxy biasing out
to $z\sim 1.5$ (Marinoni \etal~\cite{marinoni2005}).  In that study 
we showed  that, while at high redshift bright galaxies formed preferentially in
the high matter-density peaks, as the Universe ages, galaxy formation
begins to take place also in lower density environments.  
This result
on biasing evolution provides a simple and intuitive way to introduce
environment in the original downsizing picture: brighter galaxies
start forming stars earlier {\it and} preferentially in higher density
environments. 
This can explain, in a qualitative way, the
observed evolution of the colour-density relation, i.e. the faster
progressive building up of bright red galaxies in high density
environments and the slower evolution for the fainter galaxy
population.

d) The strengthening of the colour-density relation as a function of
cosmic time implies that the colour distribution has been tightly
coupled to the underlying density field at least over the past 9
Gyr: the main effect of this coupling is a marked dependence
on environment of the colour-magnitude distribution.  
The early epoch
flatness of the colour-density relation causes the bimodal colour
distribution in high density regions to mirror the one in poor
environments. However, as time goes by, the colour-density relation
strengthens and the bimodal distribution gradually develops the
present-day asymmetry between a red peak more prominent in high
density environments and a blue one mostly contributed by field
galaxies. This suggests that we have sampled the relevant time-scales over
which physical {\it nurture} processes have conspired to shape up the
present-day density dependence of the colour-magnitude distribution.

We conclude that the colour-density and colour-magnitude-density
relations are not the result of initial conditions imprinted early on
during the primordial stages of structure formation and then frozen
during subsequent evolution. Our results
suggest a scenario whereby both time evolving genetic information
(galaxy biased formation) and complex environmental interaction (star
formation quenching) concurred to build up these relations.


\end{document}